\documentclass[12pt]{article}

\newcommand{\bibi}{\bibitem}
\newcommand{\vsp}{\vskip 5mm \noindent}

\newcommand{\beq}{\vskip 5.5mm
\begin{equation}
}
\newcommand{\eeq}{
\end{equation}
\vskip 5.5mm\noindent}

\newcommand{\beqr}{\vskip 5.5mm
\begin{eqnarray}
}
\newcommand{\eeqr}{
\end{eqnarray}
\vskip 5.5mm\noindent}
\begin{document}
\newcommand{\noin}{\noindent}
\noin
{\large
{\bf
The $1/r^2$  Integrable system:
            The  Universal Hamiltonian  for  Quantum Chaos}\footnote{ To appear
in the proceedings of the
16th Taniguchi International Symposium on the Theory of
Condensed Matter: ``Correlation Effects in Low Dimensional Electron Systems''
October 23 to 29, 1993  Shima, Japan; to be published by Springer Verlag
1994 Editors N. Kawakami and A. Okiji}

\vspace{5.5mm}}

\normalsize
\noindent
B. Sriram Shastry$^\dagger$  \vspace{5mm}

\noindent
$^\dagger$ A. T. \& T. Bell Laboratories, 600 Mountain Avenue, Murray Hill, NJ
07974,
U.S.A.
\vspace{25.5mm}

\normalsize
\rm

%%%%%%%%%%%%%%%%%%%%%%%%%%%%%%%%%%%%%%%%%%%%%%%%%%%%%%%%%%%%%%%%%%%%%%
%                ABSTRACT                                            %
%%%%%%%%%%%%%%%%%%%%%%%%%%%%%%%%%%%%%%%%%%%%%%%%%%%%%%%%%%%%%%%%%%%%%%
\noindent
We summarize recent work showing that the $1/r^2$ model of
interacting particles in 1-dimension is a universal Hamiltonian
for quantum chaotic systems. The problem is  analyzed in terms
of random matrices and of the evolution of their eigenvalues
under changes of parameters.
The robustness of bulk
space-time correlations
of a many particle system to changing  boundary conditions
is suggested to be at the root of the universality.
The explicit density-density correlation functions of the $1/r^2$
model, now available through the above mapping at two values
of the coupling constant, are interpreted
in the light of Bethe's {\it Ansatz}, giving a vivid picture
of the fractionalization of bare particles or holes into ``quark''
like Bethe quasi-particles and holes.
%%%%%%%%%%%%%%%%%%%%%%%%%%%%%%%%%%%%%%%%%%%%%%%%%%%%%%%%%%%%%%%%%%%%%%
%                INTRODUCTION                                         %
%%%%%%%%%%%%%%%%%%%%%%%%%%%%%%%%%%%%%%%%%%%%%%%%%%%%%%%%%%%%%%%%%%%%%%

\vspace{10mm}
\noindent
{\bf
1. Introduction}\hfil\break
\vsp There has been a new and fascinating development
 recently, spearheaded by results from Altshuler, Simons and coworkers
\cite{bla},
which is of considerable interest to problems of quantum chaos as well as
exactly integrable many body systems. A new type of universality has
been pointed out to be operative in many quantum chaotic systems,
involving energy levels and their variation as some parameter is changed.
New and concrete results for a certain correlation function
of two variables has been computed by the use of the supersymmetry method
of Efetov \cite{efetov}, and it has been conjectured that these correlation
functions are the space time correlation functions of density fluctuators
in a very well known and well studied many body problem, the $1/r^2$ model
introduced by Sutherland\cite{bill}and Calogero\cite{calo} in the early 70's,
and indeed even earlier in a relaxational framework by Dyson during  the course
of his seminal work on random matrices\cite{dyson}. This amazing connection
between disordered systems (typifying quantum chaotic systems)
on the one hand, and exactly integrable systems on the other,
is the subject of this article. We have now a route map that shows why such
a mapping is appropriate \cite{ns,sim2} and it is my purpose to explicate the
same. We will go over the mapping in two different ways here, one\cite{ns}
quite elegant following the work of Dyson, and the other more mundane
but perhaps more instructive.
\vsp Let us consider the following  game which typifies the entire problem,
let us take two matrices , say real symmetric matrices
$H_0$ and $V$ in an $N$ dimensional space and mix them parametrically
\beq
H(\phi)= \cos(\Omega \phi) H_0 + \frac{\sin(\Omega \phi)}{\Omega} V.
\label{mix}
\eeq
The motivation for this particular form of parametrization will become
transparent later, let us just regard $\Omega$ as some number of
$O(1/\sqrt{N})$ for now,
and further $H_0$ as a diagonal matrix with  some given
eigenvalues of $O(N)$, and $V$ an off diagonal generic matrix with
 matrix elements of $O(1)$. The matrix $H(\phi)$ then is characterized by
the parameter $\phi$ and  if we should rotate the matrix
 by any orthogonal transformation, then the typical
matrix element $H_{i,j} \sim O(\sqrt{N})$, giving us eigenvalues that are
always $O(N)$, and hence their separation of $O(1)$. We can ask, how
do the eigenvalues vary as we vary the parameter $\phi$, Fig.1 shows the
result of a typical run with a $20 \times 20$ matrix.
The eigenvalues never cross:
they come close together and then move off; this is the familiar level
repulsion
at work. What is more, the picture suggests, after rotating by $\pi/2$, that we
may view these spaghetti like lines as the world lines of interacting
particles, with $\phi$ a time-like coordinate and $\lambda$ a space like
coordinate. This view happens to be completely correct, and will ultimately be
established here. Our choice of the scales was judicious, but if one
had a different set of scales for the various parameters, we could
simultaneously rescale the total $H$ and the $\phi$ to get a similar spectrum.
\vsp
The universality of Altshuler and Simons \cite{bla} is
obtained in a slight
variant of the above obtained by taking the limit $\Omega \rightarrow 0$,
and that of large $N$ giving a thermodynamic limit and
some hope of ergodicity, enabling one to define an average by taking the
mean over several energy levels. The next step involves the choice
of the matrix $V$, one can always remove the mean of $V$, and further it
is suggested that the variance of $V$ can be adjusted to be unity by
rescaling the parameter $\phi$. The independence of the final correlations
to the various choices of $V$ and $H_0$, and indeed the size of the Hilbert
space have been checked with various examples \cite{bla}.
Let us recall that universality of the Random Matrix ensembles is
a statement of the universality of the ``time independent", or ``static"
correlations in the above sense \cite{pechukas}. The connection between the
static correlations
and the ground state correlation functions
of the $1/r^2$ many body
 models are of course well known, and date back to the work
of Sutherland \cite{bill}, who pointed out that the three ensembles, Orthogonal
Unitary and Symplectic correspond to just three values of the coupling constant
 $\lambda$
namely $\lambda= 1/2,1,2$, or equivalently $\beta\equiv 2 \lambda =1,2,4$,
in the $1/r^2$ model Hamiltonian
\beq
H_S= -\sum_i \partial^2/\partial x_i^2 + 2 \lambda (\lambda-1)\; \sum_{i < j}
\frac{(\pi/L)^2}{\sin^2(\pi(x_i-x_j)/L)}. \label{billh}
\eeq
The work of Altshuler and Simons \cite{bla} then implies
that the above model
is a Universal Hamiltonian for Quantum chaos in the sense that its
time dependent density density correlation functions are also universal!
\vsp
\noin
The plan of this article is as follows, we will recall the random matrix
ensembles
in Section 2, and focus on the Orthogonal ensemble, where will
see the workings of the essentially rigorous mapping\cite{ns}
between the eigenvalue
motion problem and the Sutherland Calogero model in the thermodynamic limit.
In Section 3, we will discuss a more elementary proof of the equivalence
using simpler tools.  This derivation gives a broader equivalence
for non-gaussian $V$'s for a specific initial condition providing
us with the ideaf that universality
can be viewed as the independence of bulk correlations in a many body system to
the boundary effects.  In Sections  4.1 and 4.2, we will discuss the
results of the calculations for the correlations within the context of
the exact solution of the model by Bethe's Ansatz \cite{mucci}, and provide a
novel interpretation of the correlation functions as composite objects, made
out
of Bethe quasiparticles.
Section 4.1 summarizes the results of the Bethe's Ansatz for the
$1/r^2$ model giving {\it all } the eigenvalues of the model, and
Section 4.2 expresses the calculated correlation functions for
the Orthogonal and Symplectic ensembles
in terms of the Bethe spectrum. This is a kind of ` quark'' spectroscopy
in that the bare electrons of the $1/r^2$ system are found to
breakup into more basic constituents, ``quarks'' obeying the exclusion
principle of Pauli.

\noindent
\vskip 10mm
\noindent
{\bf 2. The Mapping for  Gaussian $V$ }\hfil\break
\vsp Let us now consider the problem\cite{ns} posed by the mixing problem of
Eq(\ref{mix}). We are interested in considering the eigenvalues of
the matrix $H$ as functionals of the matrix $V$ at a given $\phi$.
We will consider  the matrix $V$ to be a Gaussian, i.e. (suppressing an overall
normalization factor everywhere in the article)
\beq
p(V)\;dV = \exp(-\frac{1}{2} Tr(V^2)) \; \prod_{i \leq j} d V_{i,j}.
\label{gauss}
\eeq
Therefore the eigenvalues   $\mu \equiv
\{\mu_1, \mu_2, \mu_3, \ldots \}$ of $H$ are now also random variables
with a certain distribution $P(\mu ,\phi | \mu_0)$,  and $\mu_0$  the initial
eigenvalues,
i.e. those of $H_0$ which may be chosen diagonal without any loss
of generality.  We are interested in the evolution of $P$ with $\phi$,
clearly at $\phi= \pi/(2 \Omega)$, the matrix $H= \frac{1}{\Omega} V$, and
the eigenvalues distribution must equal that of $V$ apart
from a simple rescaling. Let us briefly recall that this  is easily obtained
for any distribution (i.e. Gaussian or otherwise) through the methods of
invariant integration in   group theory\cite{weyl}; we summarize the main
results below in a small digression.
We consider the resolution
of identity for an arbitrary real symmetric matrix $M$,
\beq
1= \frac{1}{N!} \int_{-\infty}^{\infty} \cdots  \int_{-\infty}^{\infty}
\prod_i d\lambda_i \int dO \; \delta_{Weyl}(O \lambda_d O^T - M) \prod_{i <j}
|\lambda_i - \lambda_j| \label{resunity}
\eeq
where $\lambda_d= \{ \lambda_1,\ldots,\lambda_N \}_{diag}$ is a diagonal
matrix,
 and the Weyl delta function \cite{weyldelta}
for a real symmetric matrix is a product over the number of distinct elements
${\nu_G}= N + N(N-1)/2$ of the usual $\delta$ functions: $\prod_{i \leq j}
\delta(M_{i,j})$.
The group integration( $\int dO \times 1 \equiv 1$ )
is over the manifold of $N(N-1)/2$ parameters of the Orthogonal group.
This identity is easy to establish by  realizing that only the
proximity of the
diagonalizing orthogonal matrix $O^*$ contributes to the group integral,
and the $\lambda$'s are forced to be the eigenvalues. Thus
 we can write $R_{i,j} \equiv (O^T O^*)_{i,j}= \delta_{i,j}
+ \epsilon A_{i,j} + O(\epsilon^2)$ whereby the
group integration can be shifted to an integration over the antisymmetric
matrix $A_{i,j}$, this matrix has precisely the correct number of independent
matrix elements. The  the delta function can be rotated
to $\delta_{Weyl}(\lambda_d
 - R M R^T)$, which reduces to    $ \prod_{i < j}  \delta(A_{i,j} (\lambda_i-
\lambda_j))$ and hence the identity. Note that a similar identity
also holds for the Unitary  and Symplectic ensembles. In the Unitary case,
the matrix M is an arbitrary complex but hermitean matrix and the
group integration is over the manifold $[U]$, consisting of the Unitary
group $U(N)$ with its N parameter  abelian subgroup of diagonal matrices
identified
with a single element \cite{weyl}, and thus with $N^2 - N$ parameters.
Clearly the total number of variables in the integration are $N^2$,
equalling the number of independent elements of $M$. For the Symplectic
ensemble the integration is over the $2 N^2 - N$ parameters
of the manifold $[USp(N)]$ consisting of the Unitary Symplectic group
of $2N \times 2N$ dimensional matrices,
with
its $2N$ dimensional abelian subgroup of diagonal matrices identified
with a single element. The number ${\nu_G} = N + \beta N(N-1)/2$ for the
various
cases with $\beta = 1,2,4$ corresponding to the Orthogonal,
Unitary and Symplectic ensembles respectively. With this little digression,
we see that given the probability distribution function
of any matrix ensemble in the form
$p(M)= f(Tr(M^m))$, with an appropriate arbitrary positive function $f$,
we can immediately infer the eigenvalue distribution of $M$ by convoluting
with identity and integrating out the Orthogonal group, yielding the familiar
result
\beq
p(\{ \lambda_1, \ldots, \lambda_N \}) = \prod_{i<j} |\lambda_i - \lambda_j|
f(\sum_i \lambda_i^m) \label{evseq}.
\eeq
\vsp We now make the elementary observation that the gaussian distribution
Eq(\ref{gauss}) for the matrix $V$ implies for the matrix $H$ of Eq(\ref{mix})
the distribution
\beq
P(H , \phi | H_0)= (\frac{\Omega}{\sin(\Omega \phi)})^{\nu_G}
\exp(-\frac{\Omega^2}{2 \sin^2(
\Omega \phi)} Tr(H - \cos(\Omega \phi) \; H_0)^2)  \label{gaussm}.
\eeq
In writing out the trace explicitly, we extensively  use the functions
and notation introduced by Dyson \cite{dyson},
$g_{i,j}=1 + \delta_{i,j}$  and further combine the pair of indices
$i,j$ into one single index $\mu \equiv (i,j)$, such that
$1 \leq \mu \leq {\nu_G}$. The noteworthy feature of the Gaussian is the
uncoupling of different $\mu's$, and we can write
\beq
P( H , \phi | H_0)= \prod_\mu [ \frac{\Omega}{\sqrt{1-q^2}}\;
\exp(-\frac{\Omega^2
(H_\mu - q H_{\mu_0})^2}{
g_\mu (1-q^2)})],
\eeq
with $q= \cos(\Omega \phi)$, and recognize immediately that this is the
solution of a Fokker Planck (F-P) equation for the uncoupled components $H_\mu$
\beq
\partial P / \partial t = \sum_\mu \frac{\partial}{\partial H_\mu}[
\; \frac{g_\mu}{2} \;\frac{\partial}{\partial H_\mu} + \Omega^2 \; H_\mu] P,
\label{fpmu}
\eeq
with the mapping  $t= -\log(\cos(\Omega \phi)) / \Omega^2$,
subject to the initial condition $P \rightarrow \delta_{Weyl} (H - H_0)$
as $t \rightarrow 0$. By the usual arguments, this system is equivalent
to a Langevin Equation
\beq
\dot{ H}_\mu = - \Omega^2 H_\mu + \zeta_\mu(t), \label{langevin}
\eeq
with a delta correlated white noise $< \zeta_\mu(t) \zeta_\nu (0)>=
g_\mu \delta_{\mu,\nu} \delta(t) $. The game now is to show that such
an equation for the matrix $H$ implies another for
the eigenvalue distribution, which contains the physics of
level repulsion. This was in fact already  proved by Dyson \cite{dyson},
in a very clever and beautiful argument, which emphasizes the underlying
 group theoretic structure. In essence, Dyson argues that we can
 establish and exploit
 the invariance of the above equations under a change of basis to great
 advantage. The change of basis corresponds to Orthogonal
 transformations on $H$ induced by an arbitrary $O$ via $H' =O H O^T$.
 In the component form this transformation gives $H'_\mu=\sum_\nu R(\mu|\nu)
 H_\nu$, with $\mu \equiv (i,j)$ and $\nu \equiv (k,l)$ so that
 \beq
 R(\mu | \nu) =(O_{i,k} O_{j,l} + O_{i,l} O_{j,k} ) /g_{k,l}.
 \eeq
 It is easy to show that $ \sum_\mu g_\mu R(\rho|\mu) \; R(\nu |\mu)=
 g_\rho \delta_{\rho,\nu}$,  as well as the fact that the inverse of $R$
 is obtainable
 through $R^{-1}(\mu|\nu) = \frac{1}{g_\nu} R(\nu | \mu) g_\mu$.
 Armed with these relations we can show the basis independence of
 the Langevin equations Eq(\ref{langevin}) easily since under a change of
 basis $\zeta_\mu'(t) = \sum_\nu R(\mu| \nu) \zeta_\nu(t)$, and hence
 the correlator $<\zeta' \zeta'>$ is unchanged. We may also show the invariance
of the F-P equation Eq(\ref{fpmu}) under the change of basis by a direct
computation.
 What is the advantage of
 showing the basis independence, one may ask. The answer given by Dyson,
 is that second order perturbation theory gives the exact result for
the  eigenvalue distribution function, provided one chooses a suitable basis.
At time $t$ we have some $H$ with certain eigenalues, and we first perform
an orthogonal transformation to diagonalize $H$. Now if we increase time
to $t + \delta t$, the matrix $H$ changes only infinitesimally away from the
diagonal with a change $\delta H_{i,j} $ satisfying  $<\delta H_{i,j}>
= - \delta t \;  \delta_{i,j} \;  \lambda_i \, \Omega^2 $,
and $<(\delta H_{i,j})^2>= \delta t \; g_{i,j}$. Second order perturbation
theory suffices to fix the change of the eigenvalues to $O(\delta t)$,
and we take the averages of the equation $\delta \lambda_i =
\delta H_{i,i} + \sum_{j \neq i} (\delta H_{i,j})^2/(\lambda_i - \lambda_j)$
and its square
to obtain a Langevin equation for eigenvalues or equivalently a F-P equation
\beq
\frac{\partial}{ \partial t} P = \sum_i \frac{\partial}{  \partial \lambda_i }[
\frac{ \partial}{\partial \lambda _i} + \lambda_i  \Omega^2 + \sum_{j \neq i}
\frac{1}{\lambda_j- \lambda_i} ] P.
\eeq
We may perform a similarity transformation followed by a Wick
rotation to get a Schroedinger equation of a many body problem. In brief,
let us denote the F-P equation by the abstract equation $\partial / \partial t
|P> = - \Pi |P>$ with $\Pi$ the F-P operator having eigenvalues $\geq 0$, and
with the coordinate space
wavefunctions being obtained through the usual projections $ P(x)=
<x| P> $, with $x$ standing for the collection of $\lambda's$. The special
vector $<0|\equiv \int \; dx <x|$ satisfies $<0| \Pi =0$  so that the
probability
is conserved in time, and hence $\Pi^\dagger |0>=0$. We seek to Hermiticize
the F-P operator through a similarity transform $S^{-1} \Pi S = H$,
where $S$ is hermitean  and invertible operator depending only on the
coordinates,
so we must have the ground state condition $ H S|0> = 0$. We define
a wavefunction through $|\psi>= S^{-1} |P>$ giving $H |\psi> = - \partial/
\partial t |\psi>$, and hence at equilibrium $ | \psi_0>= S |0>$
with the notation meant to remind us that this is the ground state wave
function of $H$, hence
$|P_{eq}> = S^2 |0>$.  The time evolution of the initial condition
may be written as $|P(t)> = S \exp (-t H) S^{-1} |P(0)>$, and we can thus
make the correspondence between correlation functions of the F-P problem
and the Quantum problem by noting that the classical time dependent
correlation function $<Q_a(t) Q_b(0)> =
<0|Q_a S \exp(-t H) S^{-1} Q_b |P(0)>$. If we choose
$|P(0)>=|P_{eq}> = S | \psi_0>$ then  provided $[Q,S]=0$, i.e.
position dependent variables, we
find $<Q_a(t) Q_b(0)>= <\psi_0| Q_a \exp(-t H) Q_b | \psi_0>$.
Similar aguments can be given for multi time correlations, and one has
a mapping between the correlations of the F-P problem and the Euclidean
Schroedinger eqn. In the present problem, we can carry out the required
similarity transformation readily and find
\beq
H_C=-\sum_i{{\partial^2}\over{\partial\lambda_i^2}}- \frac{1}{2} \sum_{i>j}
{1\over{(\lambda_i-\lambda_j)^2}}+
{\Omega^4 \over{ 4}}\sum_i \lambda_i^2.\label{calogero}
\eeq
Similar results hold for the Unitary and Symplectic cases with
different values of the interaction. This model has a ground state
that is known\cite{calo,bill} $\psi_0= \prod_{i< j} |\lambda_i- \lambda_j|
\exp(-  \frac{\Omega^2}{2 }\sum_i \lambda_i^2)$. Regarding $|\psi_0|^2$
as a probability distribution of particles, the density of particles
obeys the Wigner semi circle law, $d( \lambda)= d(0) \sqrt{1-\frac{\Omega^2
\lambda^2}{2 N} }$
with $d(0)= \frac{\sqrt{2 N}}{\pi} \Omega$. We now see that this
model has an average interparticle spacing that varies from point to point
when $\lambda$ becomes large enough
but there is a central region, where the density is $d(0)$. We expect
on thermodynamic grounds that the bulk properties of the system should be
identical to those of the Sutherland model Eq(\ref{billh}), a model with
periodic
boundary conditions  and hence no surface effects,
provided
we choose the two to have the same  densities via $d_{Calogero}(0)=
d_{Sutherland}$.  This is an example of ``universality''in the
statics, which is
recognized to be a statement of the independence of bulk
static correlations on
the shape of the boundaries that we are familiar with in
statistical mechanics. If we choose a sensible thermodynamical limit then
$d_{Calogero}(0)= O(1)$, implies $\Omega = O(1/N^{1/2})$.
We infact now set $d(0)=1$ giving $ \Omega = \pi /\sqrt{2 N}$
This requires $\lambda$ to be of $O(1)$, and the surface
effects become relevant when $\lambda$ becomes of $O(N)$,
the density remains unity for $\lambda$  of $O(1)$, and the surface
effects become relevant when $\lambda$ becomes of $O(N)$.
Using the above, write finally the
density density correlation function for the $1/r^2$ model
with $r$ as the spatial separation and
time separation $t= - \ln(\cos(\Omega \phi))/ \Omega^2$ in the form
\vsp
\beqr
<\rho(r,t) \rho(0,0)> & = & \int \, dH\; dH_0 \; P(H, t |H_0) \;
\exp(-\frac{\Omega^2}{2}
{\rm Tr} H_0^2 ) \nonumber \\
& &<{\rm Tr}\Big[\delta\big(\epsilon- r -H\big)\Big]
{\rm Tr}\Big[\delta\big(\epsilon-H_0\big)\Big] >_\epsilon
\eeqr
or using  Eq(6),
\vsp
\beqr
<\rho(r,t) \rho(0,0)>&=& \frac{\Omega^{\nu_G}}
{|\sin^{\nu_G} (\Omega \phi)|}\int d{H}\;dH_0 \;
<{\rm Tr}\Big[\delta\big(\epsilon- r -H\big)\Big]
{\rm Tr}\Big[\delta\big(\epsilon-H_0\big)\Big] >_\epsilon \nonumber \\
& &\exp\bigg\{-{\Omega^2\over{2\sin^2(\Omega \phi)}}
{\rm Tr}\big[H_0^2+H^2-2H_0 H\cos(\Omega \phi)\big]\bigg\},
\eeqr
and taking the limit $\Omega \rightarrow, \;
N \rightarrow \infty $ and averaging over
$\epsilon$ a simpler expression:
\beq
<\rho(r,t) \rho(0,0)>= \frac{1}{(2 t)^{ \frac{\nu_G}{2}}}
\int_{-\infty}^{\infty}   \frac{da}{ 2 \pi}
\int d H \; d H_0  \, Tr[e^{ - ia (  r + H)}] \;
Tr[e^{ i a H_0}] \, e^{  - \frac{1}{4 t} Tr( H - H_0)^2 }.
\eeq
\vsp Generalizations of this are given for all ensembles in \cite{ns},
and the noteworthy aspect of this formula is the absence of path integration,
we have ``merely'' a two matrix integral!
A direct evaluation of this
expression is  of interest, but the result of \cite{bla}
is presumably the answer. Similarly we may write down for a three point
function by breaking up the time propagation into two
intervals:
\vsp
\beqr
<\rho(r_2, t_2) \; \rho(r_1,t_1) \; \rho( 0,0) >&=&
\int \; dH_2 \; dH_1 \; dH_0 \;
 P(H_2, t_2-t_1)\; P(H_1, t_1 |H_0)  \;\exp(-\frac{\Omega^2}{2}
{\rm Tr} H_0^2 ) \nonumber \\
& &< {\rm Tr}\Big[\delta\big(\epsilon- r_2 -H_2\big)\Big]
{\rm Tr}\Big[\delta\big(\epsilon- r_1 -H_1\big)\Big]
{\rm Tr}\Big[\delta\big(\epsilon-H_0\big)\Big] >_\epsilon.
\eeqr

\vspace{10mm}
\noindent
{\bf 3. A Direct Method for  General distribution of  $V$ }\hfil\break
\vsp We will now consider a   generalized mixing
problem with real symmetric matrices,
\beq
H(\phi) = \alpha(\phi) H_0 + \beta(\phi) V, \label{mix2}
\eeq
where $\alpha$ and $\beta$ are some arbitrary functions of the
parameter $\phi$, and further the probability distribution (P.D.) of
$V$ is no longer Gaussian, but say
\beq
p(V) dV  = \exp ( - \frac{1}{2} Tr(V^2) - \gamma  \;
Tr(V^4)) \prod_{i \leq j} d V_{i,j}, \label{genmix}
\eeq
so that we can recover the Gaussian case by letting $\gamma \rightarrow 0$.
Other non Gaussian distributions may be treated in a  similar fashion.
The eigenvalues $\lambda_n$
of Eq(\ref{mix2}) are functionals of $V$ and we are
interested in the P. D. function
\beq
P(\{ \mu_n \} | \phi) = \int \prod_{i=1}^N\delta( \mu_i - \lambda_i[V]) \; p(V)
dV, \label{defpdf}
\eeq
and begin by writing its  $\phi$ derivative
\beq
\partial P / \partial \phi = - \sum_n \int \prod_{i \neq n} \delta (\mu_i
- \lambda_i) \; \delta'(\mu_n- \lambda_n)\;
[\frac{ \partial\lambda_n}{ \partial\phi}] \, p(V) \;dV
\eeq
with the derivative $\frac{ \partial\lambda_n}{ \partial\phi} = <n| H'(\phi)
|n>$
in the running eigenbasis of $H$, i.e. $H(\phi) |n> = \lambda_n(\phi) |n>$.
We now introduce the orthogonal transformation which diagonalizes
$H$, i.e. $\{ \lambda_i \}_{diag} = O H O^T$. The eigenvalue
equation for $H$ in component form reads $H_{i,j} \psi^n_j= \lambda_n
\psi^n_i$ and hence the orthogonal matrix has elements $O_{i,j}= \psi^i_j$.
Thus
\beq
\frac{ \partial \lambda_n}{ \partial \phi} =
2 \sum_{i \leq j} O_{n,i} O_{n,j} \frac{H'_{i,j}}{g_{i,j}} \label{lamder},
\eeq
and further we rewrite $H' = a(\phi) H + b(\phi) V$, with
$a(\phi)= \frac{\alpha'}{\alpha}$ and $b(\phi)= \frac{ \alpha \beta' -
\alpha' \beta}{\alpha}$. Hence we find
\beq
\partial P / \partial \phi = -a(\phi) \sum_n \frac{\partial}{\partial \mu_n}
(\mu_n P)
 - 2 b(\phi)
 \sum_n \int \prod_{i \neq n} \delta (\mu_i
- \lambda_i) \; \delta'(\mu_n- \lambda_n) \sum_{k \leq l} O_{n,k} O_{n,l}
\, \frac{V_{k,l}}{g_{k,l}} p(V)\, dV.
\eeq
Let us now define differentiation w.r.t. the matrix element $V_{k,l}$
of any functional of a  symmetric matrix, such as $Tr(V^m)$ through
$\frac{\partial}{ \partial V_{k,l}} \, Tr(V^m) =2 m (V^{m-1})_{k,l}/g_{k,l}$,
i.e. the two components $V_{k,l}$ and $V_{l,k}$ are {\it not} regarded
as independent variables, but identical ones. Using this definition,
and integration by parts, we verify the identity for any functional $f[V]$,
\beq
\int f[V] p(V) \; dV\; \frac{V_{k,l}}{g_{k,l}} =
\frac{1}{2} \int p(V) \; dV \frac{\partial}{\partial V_{k,l}} f[V] -
4 \;\gamma \int p(V) dV f[V] \frac{(V^3)_{k,l}}{g_{k,l}}.
\eeq
Hence
\beqr
\partial P / \partial \phi & = & -a(\phi) \sum_n \frac{\partial}{\partial
\mu_n}
(\mu_n P)
 -  b(\phi)
 \sum_n \sum_{k \leq l} \int  p(V)\, dV
 \frac{\partial}{\partial V_{k,l}}[  \prod_{i \neq n} \delta (\mu_i
- \lambda_i) \; \delta'(\mu_n- \lambda_n)  O_{n,k} O_{n,l}]
 \; \nonumber \\
  &        &  + 4 \gamma b(\phi)
 \sum_n  \int   \prod_{i \neq n} \delta (\mu_i
- \lambda_i) \; \delta'(\mu_n- \lambda_n)  \sum_{k,l} O_{n,k} O_{n,l}
(V^3)_{k,l}.   \label{pde}
\eeqr
We are therefore obliged to take the derivatives of the eigenvalues and the
orthogonal matrix $O$ w.r.t. $V_{k,l}$, these follow from perturbative
arguments and we summarize the very useful  results
\beqr
\frac{\partial \lambda_m}{\partial V_{k,l}}  & = & 2
 \frac{\beta(\phi)}{g_{k,l}} O_{m,k} O_{m,l} \\
\frac{\partial O_{n,p}}{\partial V_{k,l}} & = & -\frac{\beta(\phi)}{g_{k,l}}
\sum_{m \neq n} \frac{1}{\lambda_m - \lambda_n}O_{m,p} ( O_{m,k} O_{n,l}+
O_{m,l} O_{n,k} ).
\eeqr
An immediate consequence of these is the simple result
\beq
\sum_{k \leq l} \frac{\partial O_{n,k} O_{n,l}}{\partial V_{k,l}}
= -\frac{\beta(\phi)}{g_{k,l}}
\sum_{m \neq n} \frac{1}{\lambda_m - \lambda_n}.
\eeq
We therefore find Eq(\ref{pde})  takes the form
\beqr
\partial P / \partial \phi & =& -a(\phi) \sum_n \frac{\partial}{\partial \mu_n}
(\mu_n P) +\; \;
\beta(\phi) b(\phi) \sum_n  \frac{\partial}{\partial \mu_n}
[ \frac{\partial}{\partial \mu_n}
 + \sum_{m \neq n} \frac{1}{\mu_m - \mu_n} ] P   \nonumber \\
& & + 4 \gamma b(\phi) \sum_n  \frac{\partial}{\partial \mu_n}
 \int \prod_i \delta(\mu_i - \lambda_i) \; \sum_{k,l} O_{n,k} O_{n,l} \;
(V^3)_{k,l} \;
 p(V) dV.    \label{pde2}
 \eeqr
 We can now define a time variable $ t= \int_0^{\phi} \beta(\phi) b(\phi) d\phi
$
 and $\xi(t) = -a/b \beta$ as well as $\eta(t)= 1/ \beta^4$, in terms of which
 Eq(\ref{pde2}) becomes
\beqr
\partial P / \partial t & = &  \sum_n \frac{\partial}{\partial \mu_n}
[\frac{\partial}{\partial \mu_n} + \xi(t) \mu_n
 + \sum_{m \neq n} \frac{1}{\mu_m - \mu_n} ] P \nonumber \\
 & &  + 4 \gamma \,\eta(t) \; \sum_n \frac{\partial}{\partial \mu_n} \int
\prod_i \delta(\mu_i - \lambda_i) [O (\mu_{diag} -\alpha(t) H_0)^3 O^T]_{n,n}
\;
p(V) dV. \label{pde3}
\eeqr
The last term in Eq(\ref{pde3}) prevents one from obtaining a
closed form F-P equation  involving only the eigenvalues $\mu 's$
unless  $\gamma=0$, showing that the Gaussian
is a  very special distribution. We summarize the various functions
for the original
parametrization in Eq(\ref{mix}): $ \alpha(\phi) = \cos(\phi \Omega) ,
\; \beta(\phi) =\sin(\phi \Omega)/ \Omega ,\; a(\phi)= - \Omega \tan(\phi
\Omega ),
\; b(\phi )= 1/ \cos(\phi \Omega)$ and $ \xi(\phi)= \Omega^2 , \; \eta(\phi)
= \Omega^4/\sin^4(\phi \Omega)$, with of course $t= - \ln \cos(\phi \Omega)/
\Omega^2$.
If $\gamma \neq 0$, however, there is
one set of initial conditions, namely $H_0= c {\bf 1}$ the totally degenerate
initial conditions, where  the last term in  becomes simple;
$4 \gamma \eta(t) \sum_n \frac{\partial}{\partial \mu_n}(\mu_n-c)^3 P$.
Although the initial conditions are not ``generic'', this case
does have the advantage of showing  that the added term is
translated into a quartic potential well, and corresponds to a different
kind of a box for the many body system. In fact, dimensionally, the last
term for the parametrization of Eq(\ref{mix})
may be regarded as arising from an energy functional of
$O(\frac{\mu^4}{N^4})$ and hence {\it both} the box boundary terms,
namely the quadratic well and the `quartic well' become visible to
the eigenvalues $\mu$ when they become of $O(N)$. Therefore provided
the initial conditions correspond to a uniform distribution of
$\mu's$ ( as in the center of the Wigner semicircle), one expects
a large window of space and hence of time wherein the boundaries
are irrelevent.
We reach  ``equilibrium'', provided we have $\alpha =0$ and
$\dot{\xi}, \dot{\eta} \rightarrow 0$.
The last term does simplify for an arbitrary initial $H_0$,
we get the
distribution to be  that of a logarithmic gas of Wigner- Dyson type with
quartic and quadratic confinement.
It appears from the above that the
robustness of static {\it and } dynamic correlations to
the boundary effects
is  at the root of the universality of Altshuler
and Simons.

\noindent
\vskip 10mm
\noindent
{\bf 4.1  Asymptotic Bethe's Ansatz and Quasiparticle Energies}\hfil\break
\vsp
We now summarize the results of the asymptotic Bethe's Ansatz
\cite{bill,bill_course}, which
gives an explicit expression for the ``particle-hole'' like
excitations underlying the system. The {\it complete} excitation
%----------------------------------------------------------
%gives an explicit expression for the ``particle-hole'' like
%excitations underlying the system. The {\it complete} excitation
spectrum for the $1/r^2$ model Eq(2.) with $\beta = 2 \lambda$ can be described
in remarkably simple
terms as follows. The total energy of a state of the system is
expressible as
\beq
E = \sum_{n} \ p_{n}^{2} \ ,
\label{eq:energBethe}
\eeq
with the ``pseudo-momenta'' $p_n$ satisfying the equation,
\beq
p_{n} = k_{n} + \frac{\pi(\beta-2)}{2L}\sum_{n\neq
m}\mbox{sign}{(k_{n}-k_{m})} \ .
\label{eq:pseudomom}
\eeq
The total momentum of the state is
\beq
P = \sum_{n} \ p_{n} = \sum_n \ k_n \ .
\label{eq:momentBethe}
\eeq
The bare momenta are given by $k_{n} = 2\pi J_{n}/L$, where the
$J_{n}$'s are fermionic quantum numbers $J_1 < J_2 <J_3 \ldots < J_N$.
Note that at $\beta=2$ the interaction is turned off and we recover
the free-fermion results. The important point is that the totality of
states for the N particle sector is obtained by allowing the integers
$J_n$ to take on all values consistent with Fermi statistics, not only
for $\beta=2$, but for {\it all} $\beta \in [1, +\infty]$. The
summation in Eq.~(\ref{eq:pseudomom}) is trivial to carry out and we
find
\beq
p_n= k_n + \frac{(\beta-2)\pi}{L} \left(n- \frac{N+1}{2}\right) \ .
\label{eq:p_n}
\eeq
We can now select an arbitrary state of the system by specifying that
states $\{k_1, k_2, \ldots \}$ are occupied, i.e. by introducing the
fermionic occupation numbers $n(k_j)=0,1$, such that
\beqr
E= \sum_{n} \varepsilon(k_{n})n(k_{n}) + \sum_{n\neq m}
v(k_{n}-k_{m})n(k_{n})n(k_{m}) +
\left[\frac{\pi(\beta-2)}{2}\right]^{2} \ ,
\label{eq:energquasi}
\eeqr
with $\varepsilon(k)=k^{2}$ and $v(k)=\pi(\beta-2)|k|/2L$. For future
reference, the ground state is represented by $n_0(k_{n})=1$ for
$|k_{n}|<k_{F}$ and $n_0(k_{n})=0$ otherwise, where $k_F= \pi d$.
\vsp We now consider the excitation spectrum near the ground state, wherein
we excite a particle-hole pair in the free Fermi system and ask what
the energy of the interacting system is by including the Hartree-Fock
back flow term. From this point onwards we measure all momenta in
units of $k_F$ and energies in units of the Fermi energy. Let us
suppose that one of the particles described by
Eq.~(\ref{eq:energquasi}) has initially a momentum $k$, with $|k|<1 $;
we promote it to some state labeled by $k+q$, with $|k+q|> 1$. The
energy cost in units of the Fermi energy is equal to
\beqr
\triangle(q,k) & = & \varepsilon(k+q) - \varepsilon(k) + 2
\sum_{|k'|<1} [v(k+q-k')-v(k-k')] \nonumber \\
& = & q^{2}+ 2 k q + \frac{(\beta-2)}{2}(2|k+q|-k^{2}-1) \ ,
\label{eq:energydiff}
\eeqr
and the momentum of this state is simply $q$. This implies that we can
associate a generalized energy corresponding to a particle
$\varepsilon_{>}(k)$ (i.e. $|k| > 1$) and a hole $\varepsilon_{<}(k)$
(i.e. $|k| < 1$):
\beqr
\varepsilon_{>}(k)&=&k^2 + (\beta-2) |k| \nonumber \\
\varepsilon_{<}(k)& = &\frac{\beta}{2} k^2 +\left( \frac{\beta}{2}
-1 \right) \ ,
\label{eq:quasienergies}
\eeqr
such that
\beq
\triangle(q,k) = \varepsilon_{>}(k+q)-\varepsilon_{<}(k) \ .
\label{eq:triang}
\eeq
Note that $\varepsilon_{>}(k)$ and $\varepsilon_{<}(k)$ are continuous
and have continuous derivatives across the Fermi surface.
\vsp We will introduce in the usual way, particle operators
$A^{\dagger}(k)$ and hole operators $B^{\dagger}(k)$ with the
convention that the momenta corresponding to these are constrained by
$|k| > k_F$ for particles and $|k| \leq k_F$ for holes, with
excitation energies
\beqr
E_A(k) & = & \varepsilon_>(k)- \mu \nonumber \\
E_B(k) & = & \mu - \varepsilon_<(-k) \ ,
\label{eq:quasiph}
\eeqr
where $\mu \equiv \varepsilon_>(k_F)$ is the ``chemical potential''.
The quasi-particle quasi-hole excitation created by the operator $
A^{\dagger}(k+q) B^{\dagger}(-k)$ then has energy $E_A(k+q)+ E_B(-k)$,
which of course is equal to $\triangle(k,q)$. Having introduced the
underlying fermionic quasi-particles quasi-holes through
Eqs.~(\ref{eq:quasienergies}), we would like to see if the excitations
generated by the bare density fluctuation operator $\rho_q$ can be
expressed in terms of the latter. One of our objectives then, is to
express the excitations of the system probed by the bare density
fluctuation operator $\rho_q$ in terms of the quasi-particle
quasi-hole operators. Recall that in Landau's Fermi Liquid Theory
\cite{Landau} one expresses the bare particles $c(k)$ in a series
involving quasi-particles and quasi-holes of the form
\beq
c(k)= \sqrt{z_k} \ B^{\dagger}(-k) + \sum_{(p,l)} M[k,p,l] \
B^{\dagger}(p) B^{\dagger}(l) A^\dagger(-k-p-l) + \ldots ,
\label{eq:landau}
\eeq
where $ | k | \leq k_F$, and a similar expansion for particles, where
$z_k$ is the quasi-particle residue. The density fluctuation operator
$\rho_q = \sum_k c^\dagger(k+q)c(k)$ then has a development in terms
of $1, 2, 3, \ldots$ pairs of (quasi) particle-hole excitations. In
one dimension, we expect $z_k$ to vanish for arbitrary non-zero
interactions, and hence the particle-hole series is expected to be
such that the single pair should not appear. The expansions are
somewhat non-unique, in view of the fact that we can add an arbitrary
number of ``zero energy'' and ``zero-momentum'' particle-hole
excitations to any given scheme.

\noindent
\vskip 10mm
\noindent
{\bf 4.2 Dynamical Correlation Functions and their Interpretation }\hfil\break
\vsp  We will now recall the results of \cite{bla} for the correlations of the
two non trivial ensembles, the Orthogonal and the Symplectic cases, and
indicate their interpretation following Ref.\cite{mucci}.
We begin by writing the structure function
\beqr
S(q,\omega) & = & \frac{1}{2\pi d} \int dr \int dt \ k(r,t) \
e^{-i(qr-\omega t)} \nonumber \ .
\eeqr
where the correlation function
\beq
k(r,t) = \langle\rho(\bar{r}-r,\bar{t}+t)\rho(\bar{r},\bar{t} \
)\rangle - d^2 \ ,
\label{eq:correl2}
\eeq
with $\rho(r,t)=\sum_{i=1}^{N}\delta(r-r_{i}(t))$.
$S(q,\omega)$  has a representation in terms of
the excited states of the system:
\beq
S(q,\omega) \equiv \frac{1}{N} \sum_{\nu\neq 0}
|\langle\nu|\rho_{q}|0\rangle|^{2} \delta(\omega-E_{\nu}+E_{0}) \ ,
\label{eq:sqw}
\eeq
where $H|\nu\rangle=E_{\nu}|\nu\rangle$, and
\beq
\rho_{q}=\int dr \ \rho(r) \ e^{-iqr} = \sum_{i=1}^{N} e^{-iqr_{i}} \ .
\label{eq:rho}
\eeq
For the Orthogonal ensemble the result is \cite{bla}
\beqr
k^{o}(r,t) & = & d^{2} \int_{-1}^{1}d\lambda\int_{1}^{\infty}
d\lambda_{1} \int_{1}^{\infty} d\lambda_{2} \
\frac{(1-\lambda^{2})(\lambda_{1}\lambda_{2}-\lambda)^{2}}
{(\lambda_{1}^{2}+\lambda_{2}^{2}+\lambda^{2}-2\lambda\lambda_{1}
\lambda_{2}-1)^{2}}
\nonumber \\ & \times & \exp[-i k_F^{2}t(2\lambda_{1}^{2}
\lambda_{2}^{2} - \lambda_{1}^{2}-\lambda_{2}^{2}-\lambda^{2}+1)/2]
\cos[  k_F r(\lambda_{1}\lambda_{2}-\lambda)] \ .
\label{eq:orthogonal}
\eeqr
Taking the Fourier transform of $k^{o}(r,t)$ in space and time yields
\beqr
S^{o}(q,\omega) & = & \frac{2q^{2}}{k_{F}^{4}} \int_{1}^{\infty}
d\lambda_{1} \int_{1}^{\infty} d\lambda_{2} \
\frac{[1-(\lambda_{1}\lambda_{2}-|q|/k_{F})^{2}]}
{(\lambda_{1}^{2}+\lambda_{2}^{2}+q^{2}/k_{F}^{2}-\lambda_{1}^{2}
\lambda_{2}^{2}-1)^{2}} \nonumber \\ & \times &
\delta(\lambda_{1}^{2}+\lambda_{2}^{2}+q^{2}/k_{F}^{2} - 1 -
\lambda_{1}^{2} \lambda_{2}^{2}-2\lambda_{1}\lambda_{2}|q|/k_{F} +
2 \omega/k_{F}^{2}) \nonumber \\
& \times & \theta(\lambda_{1}\lambda_{2}-|q|/k_{F} +1) \
\theta(1-\lambda_{1}\lambda_{2}+|q|/k_{F} ) \ .
\label{eq:orthfunc}
\eeqr
\vsp Also note that in the Symplectic ensemble the  results are
\beqr
k^{s}(r,t) & = & \frac{d^{2}}{2} \int_{1}^{\infty}d\lambda
\int_{-1}^{1} d\lambda_{1} \int_{-1}^{1} d\lambda_{2} \
\frac{(\lambda^{2}-1)(\lambda-\lambda_{1}\lambda_{2})^{2}}{
(\lambda_{1}^{2} + \lambda_{2}^{2}+\lambda^{2}-2\lambda\lambda_{1}
\lambda_{2}-1)^{2}} \nonumber \\ & \times &
\exp[-4i k_F^{2}t(\lambda_{1}^{2}+\lambda_{2}^{2}+\lambda^{2}-2
\lambda_{1}^{2} \lambda_{2}^{2}-1)] \cos [ 2 k_F r(\lambda-\lambda_{1}
\lambda_{2})] \ .
\label{eq:symplectic}
\eeqr
We take the Fourier transform of $k^{s}(r,t)$ to get
\beqr
S^{s}(q,\omega) & = & \frac{q^{2}}{64k_{F}^{4}} \int_{-1}^{1}
d\lambda_{1} \int_{-1}^{1} d\lambda_{2} \
\frac{[(\lambda_{1}\lambda_{2}+|q|/2k_{F}
)^{2}-1]}{(\lambda_{1}^{2}+\lambda_{2}^{2}+q^{2}/4k_{F}^{2}-
\lambda_{1}^{2} \lambda_{2}^{2}-1)^{2}} \nonumber \\ & \times &
\delta(\lambda_{1}^{2}+\lambda_{2}^{2}+q^{2}/4k_{F}^{2} - 1 -
\lambda_{1}^{2} \lambda_{2}^{2} + \lambda_{1}\lambda_{2}|q|/k_{F}
- \omega/4k_{F}^{2}) \nonumber \\ & \times & \theta(\lambda_{1}
\lambda_{2}+|q|/2k_{F} - 1) \ .
\label{eq:sympfunc}
\eeqr
\vsp We now turn to an interpretation of these
results in the light of the Bethe excitations.
A useful change of variables  is
\beqr
u & =& \lambda_1 \lambda_2  \nonumber \\
z & =& \lambda_1 + \lambda_2 \ .
\label{eq:newvars}
\eeqr
Further defining $k=u-q$ we find the expression for the Orthogonal
ensemble:
\beq
S^{o}(q,\omega)= \frac{q^{2}}{4} \int_{ \buildrel {|k| \leq 1} \over
{k+q > 1}} \ dk \ \frac{[-\varepsilon_{<}(k)]}{[\omega - q(k+q)]^{2}}
\frac{\theta(\omega- \triangle(k,q)+(k+q-1)^{2}/2) \
\theta(\triangle(k,q)-\omega)}{\sqrt{\triangle(k,q)-\omega} \
\sqrt{\triangle(k,q)+2(k+q)-\omega }}.
\eeq
An examination of the energy conserving delta function shows
that we can  view the excited particle
state as a combination of a particle and particle-hole pair, as
follows. We write a schematic development for $k \geq 1$
\beq
c^{\dagger}(k) \sim \sum_{\frac{1+k}{2} \leq p \leq k} A^{\dagger}(p)
A^{\dagger}(k-p+1) B^{\dagger}(-1) \ ,
\label{eq:c_dag}
\eeq
The excitation energy of this complex is readily seen from
Eqs.~(\ref{eq:quasienergies},\ref{eq:quasiph}) to be $E_A(p) +
E_A(k-p+1) + E_{B}(-1)$, with $\frac{1+k}{2} \leq p \leq k$.
The density fluctuation $\rho_{q}$ is then seen to be formally a two
quasi particle-hole object: writing $c(k) \sim B^{\dagger}(-k)$, we
have
\beq
c^{\dagger}(k) c(k-q) \sim \sum_{\frac{1+k}{2} \leq p \leq k}
A^{\dagger}(p) A^{\dagger}(k-p+1) B^{\dagger}(-1) B^\dagger(q-k) \ ,
\label{eq:bubble_orth}
\eeq
 The above
scheme for the density fluctuation operator $c^{\dagger}(k)c(k-q)$ is
indicated in Fig.~2. We may therefore regard the density fluctuation
as being built up from a particular set of (non-interacting) pair
states consisting of annihilating two particles at momenta $k-q$ and
$1$, and creating a pair with total momentum $k+1$, distributed over
all possible relative momenta with appropriate form factors.
\vsp We can perform a similar analysis for the Symplectic ensemble,
we recall for the Symplectic case, the Bethe energies
$\varepsilon_{>}(p)= p^2 +2 |p|$ and $\varepsilon_{<}(p)=2 p^2 + 1$
(Eq. \ref{eq:quasienergies})). We now rewrite Eq.~(\ref{eq:sympfunc})
using the same variables as in the previous case
(Eq.~(\ref{eq:newvars})). We find the result breaks up naturally into
two pieces $S_a$ and $S_b$, with the second piece $S_b$ only arising
for $q>2$:
\beq
S^s(q,\omega) = S_a( q,\omega) + \theta(q-2) \ S_b(q,\omega ) \ ,
\label{eq:twoparts}
\eeq
We write $u=(1+k)/2$ in $S_a$ and $u=(l-1)/2$ in $S_b$, in terms of which
the result may be written as
\beqr
S_a(q,\omega) & = & \frac{q^2}{4} \int_{ \buildrel {|k| \leq 1} \over
{k+q > 1}} \ dk \frac{[\varepsilon_{>}(k+q) -3]}{[\omega -
\triangle(k,q)-(k-1)^2+q^2]^2} \nonumber \\ & \times &
\frac{\theta(\omega- \triangle(k,q)) \ \theta(\triangle(k,q) +(k-1)^2-
\omega)}{\sqrt{\omega- \triangle(k,q)} \sqrt{\omega- \triangle(k,q)
+8(1+k)}} \ .
\label{eq:symp4_a}
\eeqr
and
\beqr
S_b(q,\omega) & = & \frac{q^2}{4} \int_{ \buildrel {|l| \leq 1} \over
{l +\alpha > 1}} \ dl \ \frac{[\varepsilon_{>}(l+ \alpha) -3]}{[\omega
- \triangle(l,\alpha)-(l+1)^2+q^2]^2} \nonumber \\ & \times &
\frac{\theta(\omega - \triangle(l,\alpha )) \
\theta(\triangle(l,\alpha) +(l+1)^2 -\omega)}{\sqrt{\omega -
\triangle(l, \alpha)} \ \sqrt{\omega - \triangle(l, \alpha) +8(1-l)}}
\ .
\label{eq:symp4_b}
\eeqr
An examination of the energy conserving delta function
shows that in this expression, the excited particle is
``normal'', whereas the hole is apparently
further fractionalized into two ``quasi-holes''
and a ``drone quasiparticle'' living at the fermi surface, i.e. a schematic
decomposition
\beq
c(k) \sim \sum_{k\leq p \leq \frac{k+1}{2}} B^{\dagger}(p-k-1)
B^{\dagger}(-p) A^{\dagger}(1) \ .
\label{eq:c_a}
\eeq
The restriction on the range of $p$ is such that we avoid double
counting the pair and have a natural ordering of the two quasi-holes.
The energy of the effective hole is then $E_B(-p) + E_B(p-k-1) +
E_{A}(1)$, with the constraint $k \leq p \leq \frac{k+1}{2}$.
The operator
$\rho_{q}$ is seen to be formally a two quasi particle-hole object,
\beq
c^\dagger(k+q) c(k) \sim \sum_{k\leq p \leq \frac{k+1}{2}}
B^{\dagger}(p-k-1) B^{\dagger}(-p) A^{\dagger}(1) A^\dagger(k+q) \ .
\label{eq:bubble_a}
\eeq
This scheme for the density fluctuation is illustrated in Fig.~2.
\vspace{5mm}
\noindent
In the second piece of $S^{s}$,
 we have schematically
\beq
c(l) \sim
\sum_{\frac{l-1}{2} \leq p \leq l} B^{\dagger}(-p) A^\dagger(-1)
B^{\dagger}(1+p-l) \ .
\eeq
 Using the quasi-energies
Eqs.~(\ref{eq:quasienergies},\ref{eq:quasiph}) this complex has energy
$E_B(p) + E_B(1+p-l) + E_{A}(-1)$, with the physical constraint
$\frac{l-1}{2} \leq p \leq l$.. Owing to momentum conservation, we must
regard the creation operator $c^\dagger(l+q)$ as $A^\dagger(l+q-2)$
times a particle-hole pair with energy zero and momentum $2$, i.e.
$A^\dagger(1)B^\dagger(1)$. We may eliminate a `zero pair'
$A^{\dagger}(-1)B^{\dagger}(1)$ and thus obtain the scheme for the
density fluctuation operator $\rho_{q}$,
\beq
c^{\dagger}(l+q)c(l) \sim \sum_{\frac{l-1}{2} \leq p \leq l}
A^{\dagger}(l+q-2) A^{\dagger}(1) B^{\dagger}(-p) B^{\dagger}(1+p-l) \
{}.
\label{eq:bubble_b}
\eeq
The term $S_b$ then evidently may be regarded as a two quasi
particle-hole object, and is illustrated in Fig.~2.
Summarizing, in process ($a$), we may regard the density fluctuation
as being built up from a particular set of (non-interacting) pair
states consisting of creating two particles at momenta $1$ and $k+q$,
and destroying a pair with total momentum $k+1$, distributed over all
possible relative momenta with appropriate form factors. Likewise, in
process ($b$), we may regard the density fluctuation as being built up
from a particular set of (non-interacting) pair states consisting of
creating two particles at momenta $1$ and $l+q-2$, and destroying a
pair with total momentum $l-1$, distributed over all possible relative
momenta with appropriate form factors.
\noin
\vskip 10mm
\noin
{\bf 6. Conclusions.} \hfil\break
\vsp In this article, we have reviewed the equivalence between
the problem of eigenvalue  ``motion'' with a parameter, and
the dynamics of the $1/r^2$ many body problem, within  a random matrix
framework. The special features of the Gaussian distribution
lead to a closed Fokker Planck equation that is equivalent to the $1/r^2$
model Hamiltonian, and  non Gaussian distributions are shown to lead
to extra terms involving the ``angular variables'' of the
diagonalizers. These
however, dimensionally correspond to different
kinds of confining boxes for the system, suggesting that the universality
of the dynamics found   by Altshuler and Simons has the thermodynamic
robustness of bulk correlations to  the boundaries at its root. We
have further discussed the results of the
dynamical density correlations of the $1/r^2$
model within the context of Bethe's ansatz excitations. This
phenomenological spectroscopy reveals an  interesting
decomposition of the bare particles into constituents, which is different
for the different ensembles.

\noin
\vskip 10mm
\noin
{\bf 6. Acknowledgements.} \hfil\break
\vsp
It is a  pleasure
to thank Boris Altshuler and Onuttom Narayan  as well as
Eduardo Mucciolo, Ben Simons and Bill Sutherland
for  lively and
stimulating discussions.

%%%%%%%%%%%%%%%%%%%%%%%%%%%%%%%%%%%%%%%%%%%%%%%%%%%%%%%%%%%%%%%%%%%%%
%               REFERENCES                                          %
%%%%%%%%%%%%%%%%%%%%%%%%%%%%%%%%%%%%%%%%%%%%%%%%%%%%%%%%%%%%%%%%%%%%%

%%%%%%%%%%%%%%%%%%%%%%%%%%%%%%%%%%%%%%%%%%%%%%%%%%%%%%%%%%%%%%%%%%%%%
%               REFERENCES                                          %
%%%%%%%%%%%%%%%%%%%%%%%%%%%%%%%%%%%%%%%%%%%%%%%%%%%%%%%%%%%%%%%%%%%%%

\newpage
\noin
\vskip 10mm
\noin
{{\bf Figures}} \hfil \break
\vsp
{\bf Fig.1} A typical spaghetti of eigenvalues for the mixing problem of
Eq(\ref{mix}), where we take a pair of $20 \times 20$  matrices
and evaluate their eigenvalues as $\phi$ varies.
\vsp
\vsp
{\bf Fig.2} Two quasi-particle and quasi-hole scheme for the Orthogonal
and Symplectic ensembles. The x's denote quasi-particles and o's
the quasi-holes. The solid line is the fermi surface with $1 \geq k \geq
-1$. The Orthogonal ensemble is interpreted as having a fractionalization
of the inserted bare particle but not of the bare hole. The
Symplectic case on the other hand corresponds to the bare hole
being fractionalized. The two pieces $S_a$ and
$S_b$ correspond to the two figures (a) and (b).

\end{document}